\documentstyle[pra,aps,twocolumn,epsfig]{revtex}
\begin{document}
%\draft
\wideabs{
\title{Bose--Einstein condensation in trapped dipolar gases}
\author{L. Santos$^{1}$, G. V. Shlyapnikov$^{1,2,3}$, P. Zoller$^{1,4}$, and
M. Lewenstein$^{1}$}
\address{(1)Institut f\"ur Theoretische Physik, Universit\"at Hannover,
 D-30167 Hannover,
Germany\\ 
(2) FOM Institute for Atomic and Molecular Physics,
        Kruislaan 407, 1098 SJ Amsterdam, The Netherlands\\
(3) Russian Research Center Kurchatov Institute, Kurchatov Square, 123182
 Moscow, Russia\\  
(4) Institut f\"ur Theoretische Physik, Universit\"at Innsbruck,
 A--6020 Innsbruck, Austria
}
\maketitle
\begin{abstract}
We discuss Bose--Einstein condensation in a trapped gas of bosonic particles interacting 
dominantly  via dipole--dipole forces. We find that in this case the 
mean--field 
interparticle interaction and, hence, the stability diagram 
are governed by the trapping geometry.
Possible physical realisations include
ultracold heteronuclear molecules, or 
atoms with laser induced electric dipole moments.  
\end{abstract}
}
%\narrowtext

Bose--Einstein condensation (BEC) of trapped atomic gases\cite{BEC,Hulet} 
offers
unique possibilities to highlight a general physical problem of how the nature
and stability of a Bose--condensed state is influenced by the character of
interparticle interaction. In this respect, especially interesting are 
ultra--cold 
gases with attractive interaction between particles (scattering length $a<0$). 
As known \cite{negat}, spatially homogeneous condensates with $a<0$ 
are absolutely
unstable with regard to local collapses. The presence of the trapping
field changes the situation drastically. This has been revealed in the 
successful experiments at Rice \cite{Hulet} on BEC of magnetically 
trapped atomic
$^7$Li ($a=-14$ \AA). As found in theoretical studies \cite{negat}, if the 
number of Bose--condensed particles is sufficiently small (of order $10^3$ in the conditions 
of the Rice experiments) and the spacing between the trap levels exceeds the mean--field 
interparticle 
interaction $n_0|g|$ ($n_0$ is the condensate density, $g=4\pi\hbar^2a/M$, 
where $M$ is the atom mass), there will be a metastable 
Bose--condensed state. 
In other words, the condensate is stabilized if the negative pressure
caused by the interparticle attraction is compensated by the quantum 
pressure imposed by the trapping potential. In some sense, this is similar 
to the gas--liquid phase transition in a classical system with interparticle
attraction: The gas phase is stable as long as the thermal pressure exceeds 
the (negative) interaction--induced pressure (see \cite{Mermin}).

The recent success in creating ultra--cold molecular 
clouds \cite{doyle,molec,Heinzen} opens fascinating prospects to achieve 
quantum 
degeneracy in trapped gases of heteronuclear molecules. In a sufficiently
high electric field "freezing" their rotational motion, these molecules
interact via the dipole--dipole forces. This interaction 
is long--range 
and anisotropic (partially attractive), and there is a 
non--trivial question of achieving BEC and manipulating 
condensates in trapped gases of dipolar particles.  

Thus far, only the interaction between (small) atomic dipoles has been
included in the discussion of the condensate properties. G\'oral et al. 
\cite{goral} considered the effect of magnetic dipole interaction  
in a trapped spin--polarized atomic condensate. Magnetic dipoles are
small (of the order the Bohr magneton $\mu_B$), and even for atoms 
like Chromium
($6\mu_B$) the magnetic interactions are dominated by the Van der Waals 
forces. Nevertheless, for a relatively small scattering length $a$ the 
condensate wave function may develop novel structures reflecting the 
interplay between the two types of forces. These effects can 
be amplified by modyfing $a$, which hopefully will soon become a
standard technique \cite{modif}, and could eventually appear 
in other systems, such as polar molecules, as pointed 
out in Ref. \cite{goral}. Similar effects 
have been discussed by Yi and You \cite{you} for ground--state atoms with  
electric dipole moments induced by a high dc field (of the order of 
10$^6$V/cm). These authors have demonstrated the validity of the 
Gross--Pitaevskii equation (GPE) for this system, constructed 
the corresponding 
pseudopotential, and determined an effective scattering length. 

In this Letter we discuss BEC in a trapped gas of dipolar particles, where 
the interparticle interaction is dominated by the dipole--dipole forces. 
Possible realizations include the (electrically polarized) gas of 
heteronuclear molecules as they have large permanent electric dipoles.
We also propose a method of creating a polarized atomic 
dipolar gas by laser coupling of the atomic ground state
to an electrically polarized Rydberg state. Similarly to the condensates 
with $a<0$, dipolar condensates are unstable in the 
spatially homogeneous case, and can be stabilized by confinement in a trap. 
However, we find a striking difference from common atomic condensates: 
In the BEC regime the sign and the value of the 
dipole--dipole interaction energy in the system is strongly influenced by the
trapping geometry and, hence, the stability diagram depends crucially on the
trap anisotropy. This offers new possibilities for controlling and 
engineering macroscopic quantum states. Remarkably, for dipoles oriented 
along the axis of a cylindrical trap 
we have found a critical
value $l_*=0.4$ for the ratio of the radial to axial frequency
$l=(\omega_{\rho}/\omega_z)^{1/2}$: 
Pancake traps with $l<l_*$ mostly provide   
a repulsive mean field of the dipole--dipole interaction, and thus the
dipolar condensate in these traps will be stable at any number of particles 
$N$.
For $l>l_*$ the stability of the condensate requires $N<N_c$, where
the critical value $N_c$ at 
which the collapse occurs is determined by the condition that (on
average) the mean--field interaction is attractive and close to 
$\omega_{\rho}$. 

We consider a condensate of dipolar particles in a cylindrical harmonic trap. 
All dipoles are assumed to be oriented along the trap axis. Accordingly, 
the dipole--dipole interaction potential between two dipoles is given by 
$V_d({\bf R})=
(d^2/R^3)(1-3\cos^2{\theta})$, where $d$ is the dipole moment, ${\bf R}$ the
distance between the dipoles, and $\theta$ the angle between the vector
${\bf R}$ and the dipole axis. The dipole--dipole interaction 
is long--range, and 
one can no longer use the pseudopotential approximation 
for the mean field. Similar to \cite{goral,you},
we describe the dynamics of the condensate wave function $\psi({\bf r},t)$ by
using the time--dependent GPE  
\begin{eqnarray}
&& i\hbar\frac{\partial}{\partial t}\psi(\vec r,t)=  
\left \{ -\frac{\hbar^2}{2m}\nabla^2+\frac{m}{2}(\omega_{\rho}^2\rho^2+\omega_z^2z^2)+
\right \delimiter 0 \nonumber \\
&& \left \delimiter 0 + g|\psi(\vec r,t)|^2 
+d^2\int d\vec r\,' 
\frac{1-3\cos^2\theta}
{|\vec r-\vec r\,'|^3}
|\psi(\vec r\,',t)|^2 \right \}
\psi(\vec r,t). 
\label{GPE}
\end{eqnarray}
Here $\psi({\bf r},t)$ is normalized to the total number of condensate 
particles $N$. The third term on the rhs corresponds to the mean--field 
interaction due to (short--range) Van der Waals forces, and the last 
term to the mean field of the dipole--dipole interaction. Assuming that
the interparticle interaction is mostly related to the dipole--dipole 
forces and $d^2\gg |g|=4\pi\hbar^2|a|/M$, we omit the (Van der Waals)
term $g|\psi({\bf r},t)|^2\psi({\bf r},t)$. 

%The quantity $r_*=Md^2/\hbar^2$ plays a role of a characteristic
%radius of the dipole--dipole interaction, i.e. the wave function of the
%relative motion of a pair of dipoles is influenced by the interaction
%at interparticle distances $|{\bf r}-{\bf r}'|\alt r_*$. The main contribution 
%to the integral in the dipole--dipole term of Eq.(\ref{GPE}) comes from distances
%$|{\bf r}-{\bf r}^{\prime}|$ of order the spatial size $s$ of the
%condensate. Therefore, assuming $r_*\ll s$, we do not renormalize the dipole--dipole 
%potential in eq.(\ref{GPE}).
The wave function of the relative motion of a pair of dipoles is influenced by the 
dipole--dipole interaction at interparticle distances $|{\bf r}-{\bf r}'|\alt r_*
=Md^2/\hbar^2$. This influence is ignored in the dipole--dipole term of Eq.(\ref{GPE}),
as the main contribution to the integral comes from distances
$|{\bf r}-{\bf r}^{\prime}|$ of order the spatial size of the
condensate, which we assume to be much larger than $r_*$.

As mentioned above, in the spatially homogeneous case the dipolar condensate
is unstable. For all dipoles
parallel to each other, by using the Bogolyubov method one easily finds the 
anisotropic dispersion law
for elementary excitations:
$\varepsilon({\bf k})=[E_k^2+8\pi E_kn_0d^2(1/3-\cos^2{\theta_k})]^{1/2}$,
where $E_k=\hbar^2k^2/2M$, $n_0$ is the condensate density, and $\theta_k$
the angle between the excitation momentum ${\bf k}$ and the direction of
the dipoles. The instability is clearly seen from the fact that at small
$k$ and $\cos^2{\theta_k}>1/3$ one has imaginary excitation energies 
$\varepsilon$.  

To understand the influence of the trapping field on the behavior of the
dipolar condensate, we have numerically simulated Eq.(\ref{GPE})
for various values of the number of particles $N$, dipole moment $d$, and
the trap aspect ratio $l$. By evolving Eq.(\ref{GPE}) 
in imaginary
time, we have found the condition under which the condensate is stabilized 
by the trapping field and investigated static properties of this (metastable)
Bose--condensed state. 

For the stationary condensate the wave function 
$\psi({\bf r},t)=\psi_0({\bf r})
\exp{(-i\mu t/\hbar)}$, where $\mu$ is the chemical potential, and the lhs of
Eq.(\ref{GPE}) becomes $\mu\psi_0({\bf r})$. The important energy scales of
the problem are the trap frequencies $\omega_z$, $\omega_{\rho}$ and the
dipole--dipole interaction energy per particle, defined as
$V=(1/N)\int V_d({\bf r}-{\bf r}')\psi_0^2({\bf r})\psi_0^2({\bf r}')d{\bf r}
d{\bf r}'$.
Accordingly, the quantity $V/\hbar\omega_{\rho}$, the aspect ratio of the trap
$l$, and the (renormalized) number of particles  
$\sigma=Nr_*/a_{\rm max}$, (with $a_{\rm max}=
(\hbar/2M\omega_{\rm min})^{1/2}$ 
being the
maximal oscillator length of the trap) form the necessary set of parameters 
allowing us to determine the chemical potential and give a full description
of the behavior of a trapped dipolar condensate.
% Figure 1
\begin{figure}[ht]
\begin{center}\
\epsfxsize=4.5cm
\hspace{0mm}
\psfig{file=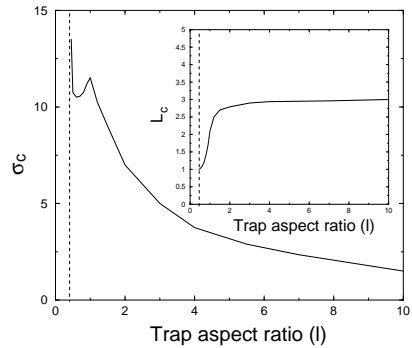,width=5.3cm}\\[0.1cm]
\end{center}
\caption{Critical value $\sigma_c=N_cr_*/a_{\rm max}$ and (in the inset) the 
corresponding condensate aspect ratio $L_c$ versus $l$.}
\label{fig:1} 
\end{figure}
We have found that the dipolar condensate is stable
either at $V>0$, or at $V<0$ with $|V|<\hbar\omega_{\rho}$. This
requires $N<N_c$, where the critical number $N_c$
depends on the trap aspect ratio $l$.
The calculated dependence $N_c(l)$ is presented in Fig.\ \ref{fig:1} and 
clearly indicates the presence of a critical point $l_*=0.4$. 
In pancake traps with $l<l_*$ the condensate is stable at any $N$, because
$V$ always remains positive (see Fig.\ \ref{fig:2}). 
For small $N$ the shape of 
the cloud is Gaussian in all directions. With increasing $N$,
the quantity $V$ increases and the cloud first becomes Thomas--Fermi in the
radial direction and then, for a very large $N$, also axially. The  
ratio of the axial to radial size of the cloud, $L=L_z/L_{\rho}$, continuously 
decreases with increasing
number of particles and reaches a limiting value at $N\rightarrow\infty$
(see Fig.\ \ref{fig:3}).  In this respect, for a very large $N$ 
we have a pancake Thomas--Fermi condensate. 

For $l\geq 1$ the mean--field dipole--dipole interaction is always attractive.
The quantity $|V|$ increases with $N$ and the shape of the cloud changes
(see Fig.\ \ref{fig:2} and Fig.\ \ref{fig:3}). In spherical traps the cloud becomes more elongated
in the axial direction and near $N=N_c$ the shape
of the cloud is close to Gaussian, with the aspect ratio 
$L=2.1$. In cigar--shaped traps ($l\gg 1$) 
especially interesting is the regime
where $\hbar\omega_z\ll |V|\ll\hbar\omega_{\rho}$. In this case the radial
shape of the cloud remains the same Gaussian as in a non--interacting gas,
but the axial behavior of the condensate will be governed by the 
dipole--dipole interaction which acquires a quasi1D character. Thus, one has
a (quasi) 1D gas with attractive interparticle interaction and
is dealing with a stable (bright) soliton--like condensate where attractive
forces are compensated by the kinetic energy \cite{soliton}. 
With increasing $N$, $L_z$ decreases. Near $N=N_c$, where
$|V|$ is close to $\hbar\omega_{\rho}$, the axial shape of the cloud also
becomes Gaussian and the aspect ratio takes the value $L\approx 3.0$.     
For $l_*\leq\l<1$, the dipole--dipole interaction energy is positive for
small number of particles and increases with $N$. The quantity $V$
reaches its maximum, and the further increase in $N$ reduces $V$ and makes
the cloud less pancake. At the critical point $N=N_c$ the shape of the cloud
is close to Gaussian and the aspect ratio $L<3.0$, tending to $1$ as
$l\rightarrow l_*$.
% Figure 2
\begin{figure}[ht]
\begin{center}\
\epsfxsize=4.5cm
\hspace{0mm}
\psfig{file=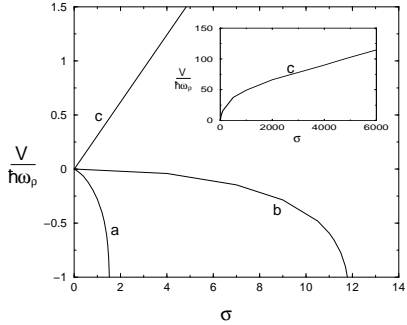,width=5.3cm}\\[0.1cm]
\end{center}
\caption{Dipole--dipole interaction energy $V$ versus $\sigma$ for 
(a) $l=10$, (b) $l=1$, and (c) $l=0.1$. In the inset the figure (c) is 
depicted in a larger scale.}  
\label{fig:2} 
\end{figure}
% Figure 3
\begin{figure}[ht]
\begin{center}\
\epsfxsize=4.5cm
\hspace{0mm}
\psfig{file=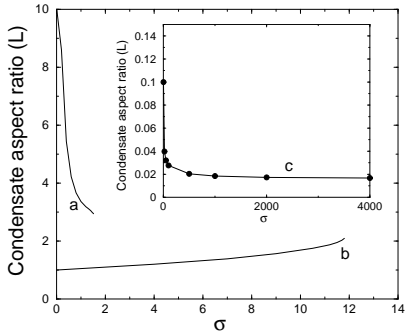,width=5.3cm}\\[0.1cm]
\end{center}
\caption{Condensate aspect ratio $L$ versus $\sigma$ for 
(a) $l=10$, (b) $l=1$, and (in the inset) (c) $l=0.1$. Figures 
(a) and (b) end when the system enters the unstable regime.}
\label{fig:3} 
\end{figure}
To gain insight in the nature of the stability of dipolar condensates in
pancake traps we performed a variational ansatz assuming a Gaussian shape
of the cloud (c.f. \cite{you}):
\begin{equation}
\label{Gaussian}
\psi_0=N^{1/2}(2\pi)^{-3/4}(L_{\rho}^2L_z)^{-1/2}
e^{-\rho^2/4L_{\rho}^2}\ e^{-z^2/4L_z^2}.
\end{equation}
Minimizing the energy functional $H$ of the system, we found the aspect ratio of 
the cloud $L$ and established that it decreases with increasing 
$\sigma$ for $l<l_{*}$, 
and increases otherwise. 
The point $l_{*}$ can be estimated 
by requiring $ dL/d\sigma |_{\sigma=0} = 0$, which provides the value
$l_*=0.41$ in good agreement with the numerical calculation.
 
For understanding the behavior of the dipolar condensate near the critical
point $N=N_c$, we observe that at this point the local minimum of $H$ 
becomes a saddle point. Hence, at $N_c$ one has 
$(\partial^2H/\partial L_z^2)
(\partial^2H/\partial L_{\rho}^2)-
(\partial^2H/\partial L_z\partial L_{\rho})^2=0$,
in addition to $\partial H/\partial L_z=\partial H/\partial L_{\rho}=0$. 
This gives the relation between $L$ and $l$ at the criticality:
\begin{equation}
\frac{(2L^2+1)(5+10l^4)}{2L^4+1} -
\frac{6B(L)(1+2l^4)}{(L^2-1)^2} -1 = 2\frac{l^4}{L^2},
\label{rela}
\end{equation}
where $B(L)=2+L^2-3L\arctan[\sqrt{1-L^2}/L]/\sqrt{1-L^2}$.
Similarly, one can find the corresponding expressions for
 $\sigma$, $L_z$ and $L_{\rho}$ as a function of $\omega_{z,\rho}$. 
The result of Eq.(\ref{rela}) differs by less than $15$\% from our numerical
calculations. 
% Figure 4
\begin{figure}[ht]
\begin{center}\
\epsfxsize=4.5cm
\hspace{0mm}
\psfig{file=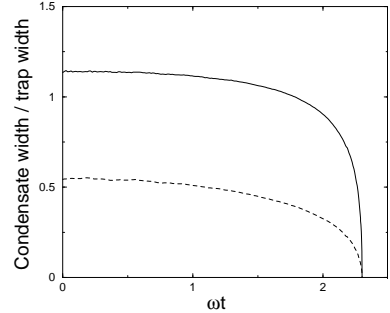,width=5.0cm}\\[0.1cm]
\end{center}
\caption{Condensate widths (in units of $a_z$)
in the dipole direction (solid line) and perpendicular to the dipole 
(dashed line), for the collapse in a spherical trap of frequency $\omega$ 
($\sigma(0)=11.5<\sigma_{c}=11.8$, and  
$\sigma(t)=11.5+0.8\omega t$).}
\label{fig:4} 
\end{figure}
We have also analyzed the dynamics of the instability, evolving (in real time)
the initially stable condensate by a slow increase of $\sigma$.
In our approach based on the GPE, similarly
to the case of $a<0$ (see \cite{Mury}), the dipolar
condensate collapses to a point on a finite time scale. The ratio 
$L$ increases moderately in the course of the collapse. The results
for the spherical trap are presented in Fig.\ \ref{fig:4}. 
The collapse in cigar--shaped
traps occurs in a similar way, since the initial shape of the collapsing
dipolar cloud is almost the same as in spherical traps 
(see Fig.1).    

As we see, the ground state of a dipolar gas exhibits a very rich behavior and one 
finds various BEC regimes. In order to electrically polarize an ultra--cold cloud of 
heteronuclear molecules, and thus create a molecular dipolar gas one should have 
the electric field which provides the Stark potential $dE$ greatly exceeding the 
spacing between the lowest rotational levels of the molecule. Then the rotational 
motion of the molecules will be "frozen" and their dipole moments will be oriented 
along the direction of the field. For most of the diatomic molecules the rotational 
level spacing is in the range from $0.1$ to $1$ K, and hence the required electric 
field is $E\sim 10^3$ V/cm. 

At present, molecular condensates are not achieved experimentally. We thus propose 
and analyze an alternative method of inducing  electric dipole moments, which can 
be used in atomic condensates. The idea is to apply a constant electric field and 
to optically admix the permanent dipole moment of a low--lying Rydberg state to the atomic 
ground state. Rydberg states of hydrogen and alkali atoms exhibit a linear Stark effect 
\cite{Gallagher}: in hydrogen, for example, an electric field $E_s$ splits the manifold 
of Rydberg states with given principal quantum number $n$ and magnetic quantum number 
$m$ into $2(n-|m|-1)$ Stark states. The outermost Stark states have (large) permanent 
dipole moments $d_R \sim n^2 e a_B$ (with $a_B$ the Bohr radius), and there will be an 
associated dipole--dipole force between the atoms. 

%The spacing $\hbar\omega_s\simeq nea_BE_s$ between adjacent Stark states should greatly 
%exceed the mean--field dipole--dipole interaction (and the gas temperature) in order to 
%avoid interaction--induced transitions from the lowest sublevel to other sublevels of the manifold. 
%
This dipole--dipole interaction can be controlled with a laser \cite{gates}. This is 
achieved either by admixing the permanent dipole moment of the Stark states to the 
atomic ground state with an off--resonant cw laser, or by a stroboscopic excitation with 
a sequence of laser pulses.  The pulses tuned to the lowest Stark state of a given Rydberg
manifold should be separated by the time $T$, have duration 
$2\Delta t\ll T$ and area $2\pi$ \cite{bambini}. 
The field $E_s$ and the "dressing" light have to be chosen such that they do not couple the 
selected lowest Stark state to other states, and the spacing ($\sim nea_BE_s$) 
between the adjacent Stark states should greatly exceed the mean--field dipole--dipole interaction 
in order to avoid the interaction--induced coupling. 
Stroboscopic excitation  ``dresses'' the atomic internal states, 
so that each atom acquires a time averaged dipole moment of the order of
$d_s=n^2ea_B f$, oriented in the direction of $E_s$. Even though the quantity $f=\Delta t/T$ 
is assumed to be small, the
induced dipole can be rather large for $n\gg 1$.  
Taking for example $\Delta t=1$ns, $T=10\mu$s, and $n=20$, we obtain
$d_s=0.1$D.
In the limit $f\rightarrow 0$, $n^2f={\rm const}$, the resulting time
dependent Hamiltonian can be replaced by its time average, leading to 
Eq.(\ref{GPE}) with $d=d_s$. A characteristic time scale in Eq.(\ref{GPE})
is of order the inverse trap frequency $\omega^{-1}$. Hence, in our case
the dynamics of the system is described by Eq.(\ref{GPE}) with
$d=d_s$, if the condition $\Delta t,T\ll\omega^{-1}$ is satisfied.
This has been tested numerically for $\Delta t/T=10^{-4}$: We reproduced our
previous results of the static GPE by solving explicitly
Eq.(\ref{GPE}) for the stroboscopic "dressing" of atoms, i.e. 
by setting
$d=d_s/f$ in the time intervals $\Delta t$ and $d=0$ otherwise.

Aside from inducing permanent electric dipoles, the stroboscopic
dressing of atoms will somewhat modify the trapping potential and the 
scattering length
related to the Van der Waals interatomic forces. 
The corresponding corrections will be proportional to the small parameter 
$f$.
There will also be losses due to spontaneous emission 
and black body radiation \cite{gates}. The rates of these processes 
become comparable with each other for $n=20$ \cite{Gallagher,gates}, 
where the corresponding decay time is of order
$20\mu$s. In our scheme the lifetime will be thus $\simeq 20\ \mu{\rm s}
/f\simeq 0.2$ s.   

The laser resonant with a bare transition frequency ``dresses'' only the atoms that are 
sufficiently separated from their neighbors, since otherwise the dipole--dipole interaction 
shifts atomic resonances. Atomic pairs are "shielded" \cite{weiner} and not dressed at 
interatomic distances smaller than $R_*$, where the latter follows from the equation 
$\hbar\Omega\simeq d_R^2/R_*^3$, with $\Omega=\pi/\Delta t$ being the Rabi frequency 
associated with the ``dressing'' laser. For $n=20$ and $\Delta t=1$ns we have 
$\Omega=500${\rm MHz} and $R_*=0.7\mu$m.
%the dipole--dipole shifts are of the order of a few GHz for interatomic 
%distances of order 1000 \AA. 

The most dangerous "underwater stone" concerns inelastic decay processes.
Fortunately, the ``shielding'' can suppress Penning ionization: 
a strong suppression is expected if atoms practically do not move
during the short time $\Delta t$, and the distances at which 
the ionization occurs ($\sim n^2 a_B$) are significantly smaller than $R_*$.
For the parameters considered above this should be the case.
The dipole--dipole interaction also induces the change of 
the effective Rabi frequency, and therefore the $2\pi$ pulse condition 
is not strictly satisfied. Hence, a fraction of atoms remains in the 
Rydberg state between the stroboscopic pulses
and decays due to spontaneous emission. This fraction can be reduced 
by decreasing the quantity $\tilde n d^2\Delta t$, 
where $\tilde n$ is the gas density.
A detailed
analysis of inelastic processes in the conditions of
stroboscopic dressing of atoms requires a separate investigation.

We acknowledge  support from   Deutsche Forschungsgemeinschaft (SFB
407), TMR ERBXTCT--96--002, the Stichting voor Fundamenteel Ondersoek der 
Materie (FOM), Alexander von Humboldt Stiftung, the Russian Foundation for 
Basic 
studies, and INTAS.
We thank K. G\'oral, H. Kn\"ockel, A. Muryshev, T. Pfau, K. Rz\c a\.zewski, 
A. Sanpera, and E. Tiemann 
for fruitful discussions.

\end{document}